# Spin-singlet superconductivity with multiple gaps in PrO$_{0.89}$F$_{0.11}$FeAs

K. Matano[1], Z.A. Ren[2], X.L. Dong[2], L.L. Sun[2], Z.X. Zhao[2] & Guo-qing Zheng[1]

[1] *Department of Physics, Okayama University, Okayama 700-8530, Japan,*

[2] *Institute of Physics, Chinese Academy of Sciences, Beijing 100190, China*

**Since the discovery of high transition-temperature ($T_c$) superconductivity in copper oxides two decades ago[1], continuous efforts have been devoted to searching for similar phenomenon in other compounds. With the exception of MgB$_2$ ($T_c$ =39 K)[2], however, $T_c$ is generally far lower than desired. Recently, breakthrough has been made in a new class of oxypnictide compounds[3-10]. Following the initial discovery of superconductivity in LaO$_{1-x}$F$_x$FeAs ($T_c$ =26 K)[3], $T_c^{onset}$ has been raised to 55 K in ReO$_{1-x}$F$_x$FeAs (Re: Ce, Pr, Nd, Sm)[5-10]. Meanwhile, unravelling the nature of the energy associated with the formation of current-carrying pairs (Cooper pairs), referred to as the 'superconducting energy gap', is the first and vital step towards understanding why the superconductivity occurs at such high temperature and is also important for finding superconductors with still higher $T_c$. Here we show that, on the basis of the nuclear magnetic resonance (NMR) measurements in PrO$_{0.89}$F$_{0.11}$FeAs ($T_c$ =45 K), the Cooper pair is in the spin-singlet state (two spins are anti-paralleled), with two energy gaps opening below $T_c$. The results strongly suggest the existence of nodes (zeros) in the gap. None of superconductors known to date has such unique gap features, although copper-oxides and MgB$_2$ share part of them.**

The most important ingredients of the superconducting gap are the symmetries of the orbital wave function and the spin configuration of the Cooper pairs, as well as the gap size. Immediately after the discovery of high-$T_c$ superconductivity in ReO$_{1-x}$F$_x$FeAs (Re: La, Ce, Pr, Nd, Sm) [3-10], many theoretical models have been put forward. These include spin-triplet, where two spins are paralleled [11-12], orbital anti-symmetric state, and spin-singlet state where two spins are anti- paralleled [13-16]. In the later case, the orbital wave function is further divided into two categories, the *d*-wave symmetry [15,16], with nodes in the gap and *s*-wave symmetry in which a gap opens everywhere on each of the Fermi surfaces [13-14]. These predictions on the symmetry stem from different assumptions for the roles of the electronic bands. Therefore, identifying the pairing symmetry will help understand the microscopic nature of the new superconductors. NMR is one of the most powerful tools to determine the symmetry of the superconducting gap. The spin susceptibility measured by the Knight shift ($K$) will vanish at a temperature far below $T_c$ in the case of spin-singlet gap, while it remains unchanged across $T_c$ for spin-triplet pairing [17]. Whether or not there exist nodes in the gap function will result in distinct temperature variation of the spin-lattice relaxation rate below $T_c$.

Figure 1 shows a typical $^{75}$As NMR spectrum in PrO$_{0.89}$F$_{0.11}$FeAs ($T_c$=45 K), which consists of a sharp central peak and two satellite peaks due to the nuclear quadrupole interaction. The width of the peak is temperature independent above $T_c$, while it increases below $T_c$ due to the formation of vortex lattice in the superconducting state. The Knight shift, which is defined as $K = (\omega/\gamma_N - H_0)/H_0$, where $\omega$ is the NMR frequency, $\gamma_N$ =7. 292 MHz/T is the nuclear gyromagnetic ratio and $H_0$ is the applied magnetic field, is shown in Fig. 2 as a function of temperature. The most important feature is that, below $T_c$, $K$ decreases down to $T$=20 K, then is followed by a still sharper drop below. The decrease of $K$ indicates spin-singlet pairing. However, such behaviour is not seen in usual spin-singlet superconductors such as copper-oxides, where $K$ decreases rapidly below $T_c$ which is followed by a milder decrease at low temperatures, as illustrated by the broken curve in Fig. 2. The underlying physics is that the gap opens rapidly below $T_c$ and then is gradually saturated to a certain value at low temperatures.

The step-wise decrease of the Knight shift is also reflected in the temperature dependence of the $^{19}$F spin-lattice relaxation rate $1/T_1$, as seen in Figure 3. There, one observes that $1/T_1$ is in proportion to $T$ above $T_c$, but drops sharply below $T_c$, which is an indication of anisotropic gap with nodes[19]. For an isotropic gap that opens everywhere on the Fermi surface, $1/T_1$ would show a coherence peak just below $T_c$ [20]. The novel feature is that, the steep drop of $1/T_1$ is gradually replaced by a slower change below $T$=33 K, and there is a broad hump-like feature around 20 K. This behaviour is also in contrast to the case of usual superconductors with a single gap, which is illustrated by the broken curve in Figure 3 for a gap with $d$-wave symmetry as an example. It should be emphasized that the uncommon temperature variation is not due to sample inhomogeneity, which would result in a two-component $1/T_1$ below $T_c$ [21]. We find that $1/T_1$ is of single component throughout the whole temperature range. Nor splitting of the spectrum below $T_c$ was found, which would occur in the case of phase separation. The data cannot be accounted for by "dirty" $d$-wave in the presence of impurity scattering, which would result in a $T_1T$=const. behaviour at low temperatures[22], as depicted by the dotted curve in Fig. 3.

We find that a two-gap model can consistently explain the step-wise temperature variation of both $K$ and $1/T_1$. The underlying physics is that the physical quantities just below $T_c$ are dominantly governed by a larger gap while the system does not 'notice' the existence of a smaller gap. Only at low temperatures where the thermal energy becomes comparable or smaller than the smaller gap, the system realizes the smaller gap, resulting in another drop of $K$ and $1/T_1$. In the superconducting state, the Knight shift ($K_s$) and the relaxation rate ($1/T_{1s}$) are expressed as [23],

$$\frac{K_s}{K_N} = \int N_s(E) \frac{\partial f(E)}{\partial E} dE \quad (1)$$

$$\frac{T_{1N}}{T_{1s}} = \frac{2}{k_B T} \iint N_s(E) N_s(E') f(E)[1-f(E')]\delta(E-E') dE dE' \quad (2)$$

Where $N_s = \frac{E}{\sqrt{E^2 - \Delta^2}}$ is the density of state (DOS) in the superconducting state, and $f(E)$ is the Fermi distribution function. By assuming $d$-wave gap symmetry $\Delta(\varphi) = \Delta_0 \cos(2\varphi)$ with the mean-field temperature dependence and that the total gap $\Delta$ is composed of two components, $\Delta = \alpha \Delta_1 + (1-\alpha)\Delta_2$, the data can be fitted reasonably well, with $\Delta_1(T=0) = 3.5 k_B T_c$, $\Delta_2(T=0) = 1.1 k_B T_c$ and $\alpha = 0.4$ (see curves in Fig. 2 and Fig. 3).

The multiple gaps likely originated from the multiple bands[24-27] of this family of compounds which consist of stacking ReO(F) and FeAs layers. First-principle calculation indicates that the Fermi surface (FS) of LaOFeAs is quasi-two dimensional and consists of hole-type sheets around the $\Gamma$ point and electron-type sheets around the M point of the Brillouin zone[24-26]. It is conceivable that the two gaps open up on the different sheets of the FS. In this context, the present compound resembles the situation of $MgB_2$ where a large gap opens on the FS derived from the orbitals in the boron plane which couples strongly to specific phonon mode, while a small gap opens on the FS derived from orbitals perpendicular to the boron plane[28]. However, unlike $MgB_2$ where the gaps open everywhere, there are zeros in the gap function of $PrO_{0.89}F_{0.11}FeAs$ as evidenced by the lack of the coherence peak and the $T^3$-like variation of $1/T_1$ just below $T_c$. This latter feature is shared by the high-$T_c$ copper oxides[29]. Therefore, the new Fe-based oxides bridge between two known classes of high-$T_c$ superconductors, $MgB_2$ and copper oxides, and the superconductivity here appears to be closely linked to its multiple electronic bands.

**Methods**

The poly-crystal of $PrO_{0.89}F_{0.11}FeAs$ was synthesized by high-pressure method[6]. Fine powders of PrAs (pre-synthesized by Pr pieces and As powder), Fe, $Fe_2O_3$, $FeF_3$ with purities better than 99.99% were mixed together according to the stoichiometric ratio, then ground thoroughly and pressed into small pellets. The pellets were sintered in a high pressure synthesis apparatus under a pressure of 6 GPa and temperature of 1250°C for 2 hours. The sample was characterized by powder X-ray diffraction (XRD) method with Cu-Ka radiation to be a tetragonal ZrCuSiAs-type structure (P4/nmm space group) [6]. For NMR measurements, the pellet was crushed into fine powders. AC susceptibility measurement using the NMR coil indicates that $T_c$ for the fine powder sample is 45 K at zero magnetic field. The powder sample was aligned in a magnetic field of $H$=9 T and fixed by an epoxy. $T_c$ is 40 K and 38 K at $H$=1.375 T and 7.5 T, respectively. NMR measurements were carried out by using a phase-coherent spectrometer. The spectra were taken at a fixed frequency of 55.1 MHz by changing the magnetic field step by step and recording the echo intensity. The spin-lattice relaxation rate was measured by using a single saturation pulse. Inspection of the crystal structure shows that the principle axis of the NQR (nuclear quadruple resonance) frequency $\nu_Q$ (found to be 12 MHz) is along the c-axis. The $\nu_Q$ tensor obtained from Fig. 1 is 6 MHz, which indicates that the obtained NMR spectrum is the *ab*-plane aligned spectrum. The Knight shift consists of three contributions, the contribution due to spin susceptibility, $K_s$, that due to orbital susceptibility, $K_{orb}$, and that due to nuclear quadruple interaction, $K_{NQI}$. The last part $K_{NQI} = 3\nu_Q^2 / 16(\gamma_N H_0)^2$ was subtracted from the raw data to give $K$ plotted in Fig. 2.

**Acknowledgements** We thank Z. Fang, S. Kawasaki, K. Kuroki, and K. Machida for useful discussion. This work was supported in part by research grants from MEXT and JSPS, Japan, NSF of China, and the ITSNEM program of CAS. The authors declare no competing financial interests.

Correspondence and requests for materials should be addressed to G.-q.Z (zheng@psun.phys.okayama-u.ac.jp).

Figure legends.

Figure 1: $^{75}$As-NMR spectrum at 55.1 MHz and $T$=40 K. The right inset shows the full width at the half maximum (FWHM) of the central transition peak as a function of temperature. The left inset compares the central transition peak at $T$= 40K (above $T_c(H)$) and 10 K (below $T_c$).

Figure 2: The temperature variation of $^{75}$As Knight shift. The solid curve is a fitting of two gaps with $\Delta_1 = 3.5 k_B T_c$ and a relative weight of 0.4, and $\Delta_2 = 1.1 k_B T_c$ with a relative weight of 0.6. The broken curve below $T_c$ is a simulation for the larger gap alone. In both cases, $K_{orb}$ was taken as 0.008%.

Figure 3: The temperature dependence of $^{19}$F spin-lattice relaxation rate ($1/T_1$) in PrO$_{0.89}$F$_{0.11}$FeAs measured at $H_0$=1.375 T. The broken line indicates a relation of $T_1T$=constant which holds for a weakly correlated electron system. The value of $1/T_1$ in the normal state is very close to that in LaO$_{0.89}$F$_{0.11}$FeAs [18]. The solid curve is a two-gap fit with the same parameters as in Fig. 2. The broken curve below $T_c$ is a simulation for the larger gap alone, and the dotted curve is for the case of impurity scattering with 50% DOS remained at the Fermi level. The thin straight line illustrates the temperature dependence of $T^3$.

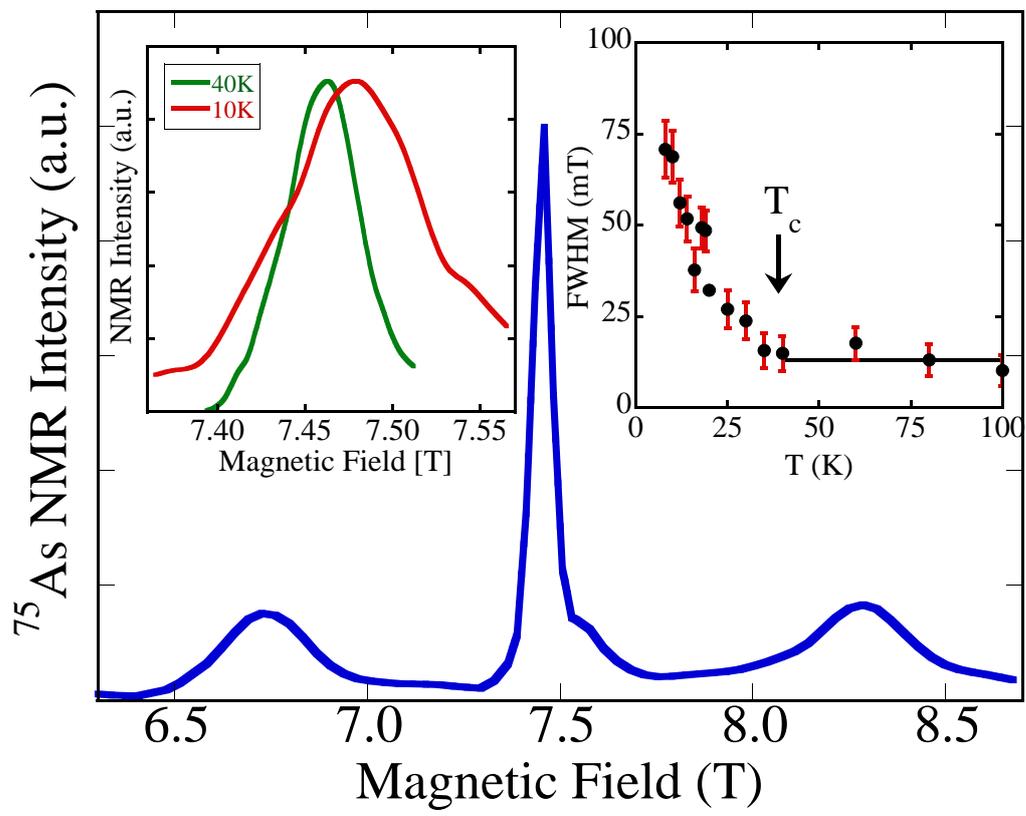

Figure 1

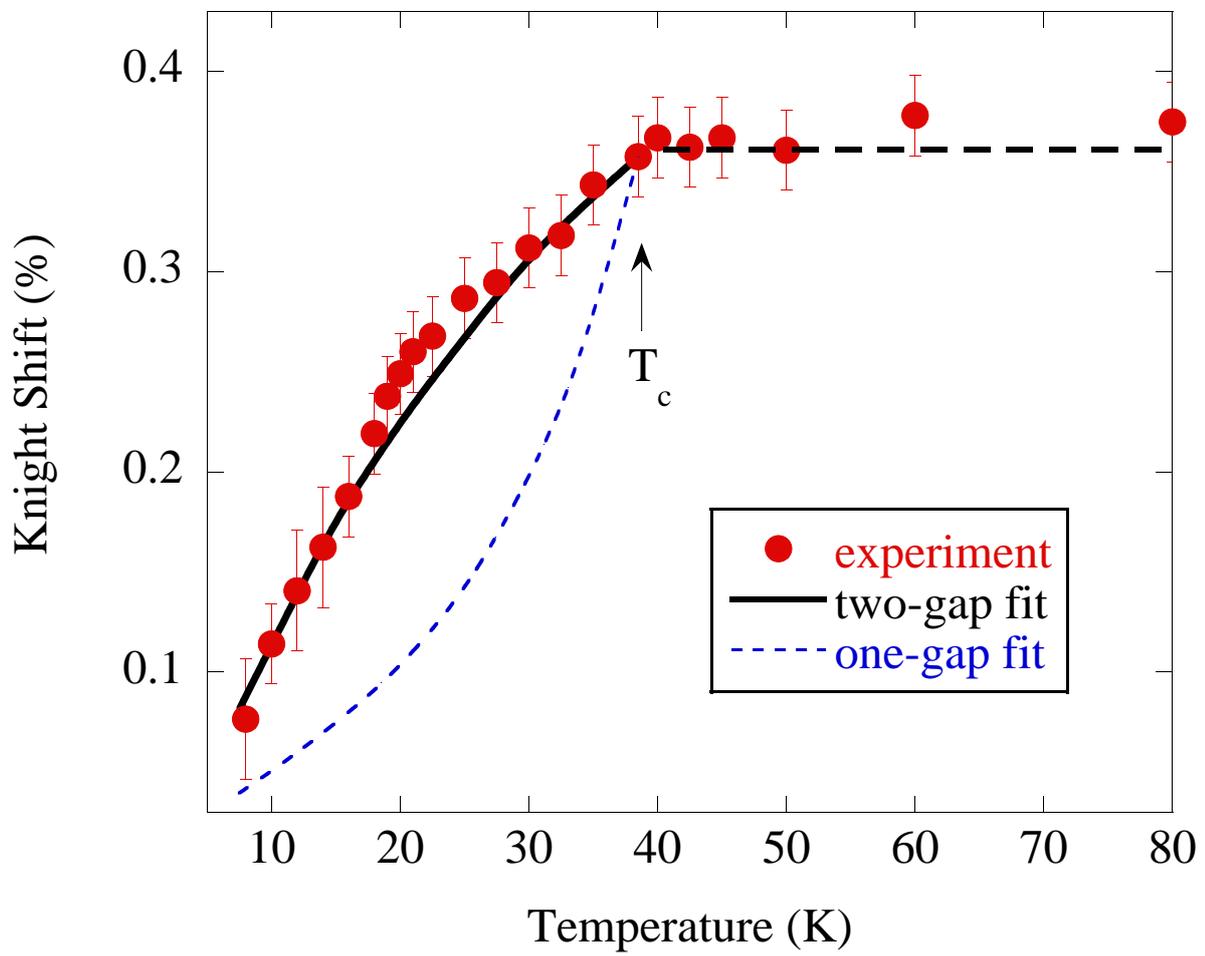

Figure 2

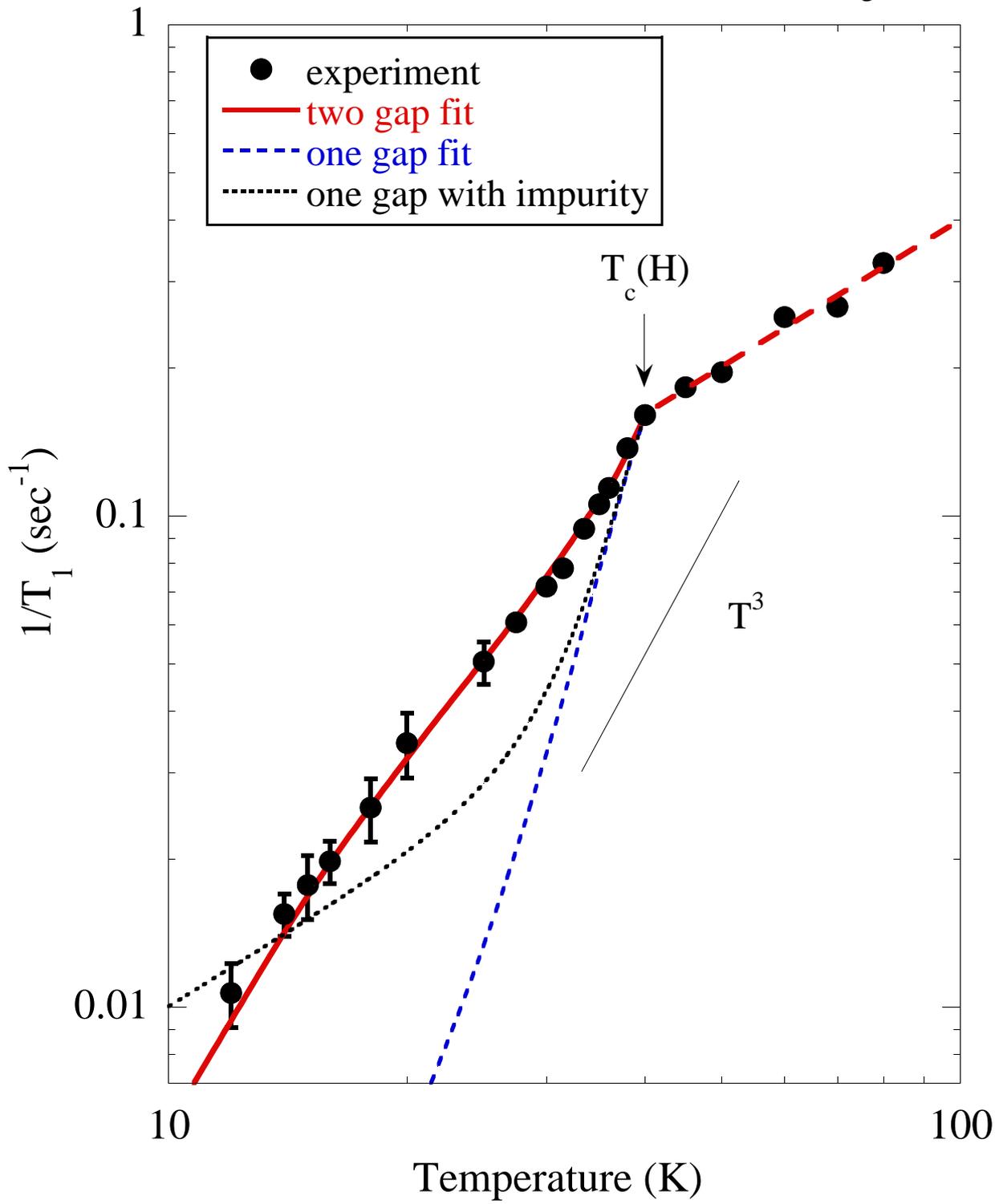

Figure 3